\documentclass[%
 reprint,
 superscriptaddress,
 amsmath,amssymb,
 aps,
prb,
prstper,
floatfix,
longbibliography 
]{revtex4-2}

\usepackage[pdftex]{graphicx} \graphicspath{{}}
\usepackage{tikz}
\usetikzlibrary{decorations.pathreplacing}
\usepackage{float}
\usepackage{dcolumn}
\usepackage{bm}
\usepackage[pdftex,colorlinks=true]{hyperref}
\hypersetup{
  colorlinks=true,
  linkcolor=blue,
  urlcolor=cyan,
}

\newcommand{\eqn}[1]{\begin{equation} #1 \end{equation}}
\newcommand{\eqa}[1]{\begin{align} #1 \end{align}}
\def\be{\begin{equation}}
\def\ee{\end{equation}}

\newcommand{\tik}[1]{\begin{tikzpicture} #1 \end{tikzpicture}}

\usepackage{color}
\definecolor{ForestGreen}{RGB}{34, 139, 34}

\newcommand{\nn}{\nonumber}

\newcommand{\mD}{\mathcal{D}}
\newcommand{\mZ}{\mathcal{Z}}
\newcommand{\mS}{\mathcal{S}}

\newcommand{\avg}[1]{\left\langle #1 \right\rangle}
\newcommand{\pd}{\partial}

\newcommand{\mO}{\mathcal{O}}

\newcommand{\mC}{\mathcal{C}}

\newcommand{\mG}{\mathcal{G}}

\newcommand{\hS}{\hat{S}}

\newcommand{\bsigma}{\boldsymbol{\sigma}}

\newcommand{\hbm}{\hat{\boldsymbol{m}}}
\newcommand{\hbS}{\hat{\boldsymbol{S}}}

\newcommand{\su}{\hat{\mathfrak{su}}}

\DeclareMathOperator{\tr}{tr}
\DeclareMathOperator{\WZ}{WZ}

\newcommand{\sectionn}[1]{\textit{#1---}}

\begin{document}

\preprint{APS/123-QED}
 
\title{Transition between Critical Antiferromagnetic Phases in the $J_1$--$J_2$ Spin Chain}

\author{Adam J. McRoberts}
\affiliation{International Centre for Theoretical Physics, Strada Costiera 11, 34151, Trieste, Italy}

\author{Chris Hooley}
\affiliation{Centre for Fluid and Complex Systems, Coventry University, Coventry, CV1 5FB, United Kingdom}

\author{A.~G. Green}
\affiliation{London Centre for Nanotechnology, University College London, Gordon St., London, WC1H 0AH, United Kingdom}

 
\date{\today}

\begin{abstract}
\noindent
The $J_1$--$J_2$ spin chain is one of the canonical models of quantum magnetism, and has long been known to host a critical antiferromagnetic phase with power-law decay of spin correlations. We show in this Letter that there are, in fact, \textit{two} distinct critical antiferromagnetic phases, where the roles of the local dimer field and its dual field are interchanged: the `Affleck-Haldane' phase near the Heisenberg point $J_2 = 0$, where the dimer field that parametrises local singlet order is gapless and part of a joint $O(4)$ N\'eel-singlet order parameter; and the `Zirnbauer' phase which appears at sufficiently large ferromagnetic $J_2$, where the dimer field is gapped out and its \textit{dual} field -- the instanton density of the $O(3)$ N\'eel field -- is critical instead.
The phases are so-named because each realises one of the competing pictures for how the $O(3)$ non-linear sigma model with a topological theta term renormalises to the $\mathfrak{\hat{su}}(2)_1$ Wess-Zumino-Witten model. 
We support these predictions with density matrix renormalisation group calculations.
\end{abstract}

\maketitle

\noindent
%
The $J_1$--$J_2$ spin-$\frac{1}{2}$ chain, the Hamiltonian of which is
\eqa{
\hat{H} = \sum_i \left(J_1 \hbS_i \cdot \hbS_{i+1} + J_2 \hbS_i \cdot \hbS_{i+2} \right),
\label{eq:Hamiltonian}
}
has, for more than a half-century, served to elucidate such questions as the role of topological terms~\cite{haldane1983continuum,haldane1985theta,haldane1988nonlinear,affleck1987critical}, competing interactions and frustration~\cite{Haldane:1982tu,OKAMOTO1992433,Nomura_1994,furukawa2012ground}, and incommensurate order~\cite{white1996dimerization,nersesyan1998incommensurate}. 

In this Letter, we will be concerned with the zero-temperature quantum phase diagram of Eq.~\eqref{eq:Hamiltonian}, and, in particular, with the antiferromagnetic chain ${J_1 > 0}$ (and ${J_2 \leq J_1/2}$); from this point, we will work in units where ${J_1 = 1}$. The central result is that there are \textit{two} distinct antiferromagnetic critical phases, cf. Fig.~\ref{fig:phase_diagram}: the `Affleck-Haldane' phase near the Heisenberg point $J_2 = 0$, where the local dimer field is gapless; and the `Zirnbauer' phase which appears at sufficiently large \textit{ferromagnetic} $J_2 < 0$, where the local dimer field is gapped out, but its dual field -- the instantons of the N\'eel order -- is critical instead.

But we first need to set the stage. In Ref.~\cite{azad2024generalised}, we showed how the low-energy field theory of Eq.~\eqref{eq:Hamiltonian} can be constructed by building the path integral from an overcomplete basis of matrix product states (MPS). The MPS path integral~\cite{green2016feynman} microscopically constructs the $O(4)$ joint-order parameter ${u = (\sin\!\chi, \cos\!\chi\,\hbm) \in S^3}$, where $\hbm \in S^2$,
\eqa{
(-1)^i\langle\hbS_i\rangle \sim \cos\!\chi \,\hbm
\label{eq:Neel_order}
}
is the N\'eel order, and
\eqa{
(-1)^i\langle \hS_i^+ \hS_{i+1}^- \!+\! \hS_i^- \hS_{i+1}^+ \!-\! \hS_{i-1}^+ \hS_{i}^- \!-\! \hS_{i-1}^- \hS_{i}^+\rangle \sim \sin\!\chi
\label{eq:singlet_order}
}
is the dimer (spin singlet) order. The field theory we obtain represents the competing N\'eel and dimer orders on an equal footing, as an $O(4)$ non-linear sigma model (NLSM) plus a Wess-Zumino (WZ) term $\Omega$ (which we will together refer to as $O(4)_{\WZ}$), perturbed by an explicit dimerisation potential. That is,
\eqa{
\mS &= i\Omega + \!\int d\tau dx \left(\gamma (\pd_\mu u)^2 - \frac{J_2}{4}\sin^2\!\chi\right), \nn \\[0.1cm]
\Omega &= \int d\tau dx\, \frac{\pi + 2\chi + \sin2\chi}{4\pi}\,\hbm\cdot(\pd_\tau\hbm\times\pd_x\hbm)
\label{eq:J1J2_action}
}
(and ${\mZ = \int\mD u \,e^{-\mS[u]}}$), where ${\gamma = \sqrt{(1-2J_2)/32}}$.
\vspace{0.2cm}

This provides an appealing microscopic construction of the field theory, though we do point out the above form can also be obtained by non-abelian bosonisation~\cite{witten1984non,gogolin2004bosonization,tsvelik2007quantum}.

\begin{figure}
    \centering

    \begin{tikzpicture}
        \draw[dashed, line width = 1.25] (0.2,0.2) -- (0+0.5,0.5);
        \draw[line width = 1.25] (0+0.5,0.5) -- (0+1,0);
        \foreach \x in {1,2,3,4,5,6} {
            \draw[line width = 1.25] (\x,0) -- (\x+0.5,0.5) -- (\x+1,0);
        }
    
        \foreach \x in {1,2,3,4,5,6} {
            \draw[gray] (\x,0) -- (\x+1,0);
            \draw[gray] (\x-0.5,0.5) -- (\x+0.5,0.5);
        }
        \draw[gray, dashed] (0.5,0) -- (1,0);
        \draw[gray, dashed] (0,0.5) -- (0.5,0.5);
        \draw[gray, dashed] (6.5,0.5) -- (7,0.5);
        \draw[gray, dashed] (7,0) -- (7.5,0);
    
        \draw [line width = 1.25] (1.8,-0.5) -- (2.7,-0.5);
        \draw (2.7, -0.5) node[right] {$J_1 > 0$, };
        
        \draw [gray] (4.3,-0.5) -- (5.2,-0.5);
        \draw (5.2, -0.5) node[right] {$J_2$};
    
        \draw[dashed, line width = 1.25] (7+0.3,0.3) -- (7,0);
        \foreach \x in {1,2,3,4,5,6,7} {
            \fill[black] (\x,0) circle (2pt);
            \fill[black] (\x-0.5,0.5) circle (2pt);
        }
    
        \draw (-0.8, 0.2) node[above] {(a)};
        
    \end{tikzpicture}
    
    \begin{tikzpicture}
        \draw[->] (0,0) -- (7.5,0) node[right] {$J_2$};

        \begin{scope}
        \clip (0.0,0.0) rectangle(2.7,0.5);
        \draw[decorate, decoration={brace, amplitude=12pt}, blue] (-0.4,0) -- (2.7,0);
        \end{scope}
        \draw[decorate, decoration={brace, amplitude=12pt}, ForestGreen] (2.7,0) -- (5.8,0);
        \begin{scope}
        \clip (5.8,0.0) rectangle (7.5,0.5);
        \draw[decorate, decoration={brace, amplitude=12pt}, red] (5.8,0) -- (7.8,0);
        \end{scope}
        
        \draw (2.7,-0.1) node[below] {$J_2^*\approx-1.3$};
        \draw (4.2,-0.1) node[below] {$0$};
        \draw (5.8,-0.1) node[below] {$J_2^d\approx0.24$};
        
        \draw (4.2,0.5) node[above] {Affleck-Haldane};
        \draw (1.2,0.5) node[above] {Zirnbauer};
        \draw (6.75,0.5) node[above] {Dimerised};


        \fill[black] (2.7,0) circle (2pt);
        \fill[black] (5.8,0) circle (2pt);

        \draw (0, 1.1) node[above] {(b)};
        
    \end{tikzpicture}
    \caption{(a) Sketch of the ${J_1{\text{--}}J_2}$ spin chain, showing the nearest-neighbour $J_1$ bonds (black, thick) and next-nearest-neighbour $J_2$ bonds (grey, thin). (b) Sketch of the phase diagram for $J_1 > 0$. The critical Affleck-Haldane phase around $J_2 = 0$ is described by the $O(4)$ NLSM with a Wess-Zumino term ($O(4)_{\mathrm{WZ}}$), where the $O(4)$ field comprises both local N\'eel and dimer order (cf. Eqs.~\eqref{eq:Neel_order} \& \eqref{eq:singlet_order}). At $J_2^* \approx -1.3$ the model transitions to the critical Zirnbauer phase: the dimer field acquires a mass and decouples from the joint-order parameter, and its dual field -- the instanton density of the N\'eel field -- is gapless instead. The critical part of both phases is the same conformal field theory in the infrared ($\su(2)_1$) but the microscopic nature of the critical field changes and can be detected in the dimer correlation functions, cf. Fig.~\ref{fig:DMRG}.
    }
    \label{fig:phase_diagram}
\end{figure}

One advantage of this formulation of the field theory (as opposed to abelian bosonisation, or the $O(3)$ version discussed shortly) is that the spontaneous dimerisation transition at ${J_2^d \approx +0.24}$~\cite{Haldane:1982tu,OKAMOTO1992433,Nomura_1994} is very straightforward. For $J_2 > J_2^d$, the potential term $-J_2\sin^2\!\chi$ is relevant, and $\chi$ simply acquires a non-zero expectation value. On the other hand, the potential term is irrelevant for $J_2 \approx 0$, and we explicitly recover the correct critical field theory -- the $\su(2)_1$ Wess-Zumino-Witten model, expressed as $O(4)_{\WZ}$. Both the dimer and N\'eel correlations have power-law decay (with the same exponent), and the $SU(2)$-matrix primary field of $\su(2)_1$ can be expressed as
\eqa{
\hat{g} = \sin\!\chi\,\mathbf{1} + i \cos\!\chi\,\bsigma\cdot\hbm,
\label{eq:matrix_field}
}
where $\bsigma$ is the vector of Pauli matrices.

This Letter, however, focuses on the case where $J_2 < 0$ is \textit{ferromagnetic}, to which much less attention has been given. Of course, the interactions are not magnetically frustrated in this regime; but the ferromagnetic second-neighbour interactions do suppress the formation of local spin singlets -- one might wonder, then, whether there is some critical value $J_2^* < 0$ below which the dimer field $\chi$ in Eq.~\eqref{eq:J1J2_action} is gapped out. 

A first serious attack on this question, however, would seem to suggest no such phase transition could occur. To explain why, we need to consider an alternative description of the $J_1$--$J_2$ chain \eqref{eq:Hamiltonian} as an $O(3)$ NLSM with a topological theta term (which we will refer to as $O(3)_\pi$),
\eqa{
\mS &= i\Theta(\pi) + \gamma \int d\tau dx\, (\pd_\mu\hbm)^2, \nn \\
\Theta(\theta) &= \int d\tau dx\, \frac{\theta}{4\pi}\,\hbm\cdot(\pd_\tau\hbm\times\pd_x\hbm) \in \theta\;\!\mathbb{Z},
\label{eq:O3NLSM}
}
which is derived by analysing Eq.~\eqref{eq:Hamiltonian} with the spin coherent state path integral and applying the celebrated Haldane map~\cite{haldane1983continuum,haldane1985theta,haldane1988nonlinear,auerbach2012interacting}.

At a glance, it seems that $O(3)_\pi$ does \textit{not} describe the spin chain near ${J_2 = 0}$, since Eq.~\eqref{eq:O3NLSM} is just Eq.~\eqref{eq:J1J2_action} with the dimer field $\chi \to 0$ frozen out, and so would appear to miss the power-law dimer correlations. However, under renormalisation semi-group (RG) flow, ${O(3)_\pi \to \su(2)_1}$~\cite{lieb1961two,polyakov1984goldstone,affleck1986proof,zamolodchikov1992massless,haldane1983continuum,witten1984non,haldane1985theta,affleck1987critical,haldane1988nonlinear,affleck1988critical,senthil2006competing,auerbach2012interacting,francesco2012conformal,baxter2016exactly}, so there is an emergent operator $\sim \tr(\hat{g})$ that is typically identified with the dimer field and, indeed, has algebraic decay. 

The standard picture for how this happens is due to Affleck and Haldane~\cite{affleck1985critical,affleck1987critical}, and is summarised in Ref.~\cite{zirnbauer2024infrared}. In brief, `weak-coupling' (large $\gamma$) perturbation theory shows that $\gamma$ decreases towards the infrared -- that is, we flow towards strong-coupling. But as we enter the strong-coupling regime, $\theta$ and $\gamma$ should no longer be thought of as homogeneous in spacetime and develop correlated fluctuations, because the large variations in $\hbm$ which occur in this regime cannot be absorbed in a single RG step whilst holding $\theta$ and $\gamma$ constant~\cite{zirnbauer2024infrared}.
This motivates the introduction of an extension field $\chi(\tau, x)$ to account for these correlated fluctuations, which leads to
\eqa{
\mS = i\Theta(\theta(\chi)) + \frac{1}{4\pi}\int d\tau dx \bigl[(\pd_\mu\chi)^2 + \gamma(\chi)(\pd_\mu\hbm)^2 \nn \\
+ M^2\sin^2\!\chi\bigr].
\label{eq:ext_field_action}
}
If, under the RG flow, we find
\eqa{
\gamma(\chi) \to \cos^2\!\chi,\;\; \theta(\chi) \to \pi + 2\chi + \sin2\chi,\;\; M^2 \to 0, \nn
}
then we explicitly recover $O(4)_{\WZ}$, and we identify the extension field $\chi$ with the local dimer field; indeed, the $\chi$ fields that appear in Eqs.~\eqref{eq:J1J2_action} \& \eqref{eq:ext_field_action} are functionally equivalent. Now, if this is the only way by which $O(3)_\pi$ can become $\su(2)_1$ in the infrared, then the local dimer field in Eq.~\eqref{eq:J1J2_action} will always become massless, and there is no transition at any $J_2 < 0$.

In a recent pre-print~\cite{zirnbauer2024infrared}, however, Zirnbauer has raised some objections to this Affleck-Haldane picture, namely: (i) that the necessary masslessness of the extension field is not protected by any symmetry; (ii) although the mass term is an irrelevant perturbation when $M^2$ is small~\cite{affleck1987critical}, its bare value is infinity; and (iii) there is no guarantee that the RG flow will induce the Wess-Zumino (WZ) term required to offset the curvature of $S^3$ (a necessary condition for a fixed point~\cite{friedan1985nonlinear,braaten1985torsion}).

In Zirnbauer's alternative picture, the extension field is an artefact of the early stages of the RG flow and is frozen out in the infrared. The $\su(2)_1$ fixed point is instead reached when the target space geometry flows from $S^2$ to the parallelisable $S^1\times\mathbb{R}$ with a complex metric tensor, and the theory is dual to a free boson compactified at the radius $r = 1/\sqrt{2}$, which is dual to $\su(2)_1$~\cite{ginsparg1988curiosities}. With respect to the $J_1$--$J_2$ spin chain, this provides a plausible scenario for a different critical phase where $\chi$ is gapped. 

Armed with this idea, then, we return to the lists for another tilt at the question of whether there are, in fact, \textit{two} critical phases in the $J_1$--$J_2$ spin chain, cf.\ Fig.~\ref{fig:phase_diagram}: the `Affleck-Haldane' phase for $J_2^* < J_2 < J_2^d$, where the local dimer field $\chi$ is gapless and part of the joint $O(4)$ order parameter; and the `Zirnbauer' phase for $J_2 < J_2^*$, where the local dimer field is gapped out, and, as will be discussed a little later, its \textit{dual} field -- the \textit{instantons} of the N\'eel field $\hbm$ -- has power-law correlations instead. 

In the rest of this Letter, we first ask whether the Zirnbauer phase is perturbatively stable (at least, self-consistently) under one-loop RG, and find there is a finite critical bare mass $M_*^2$ above which it is stable -- and below which it is unstable -- which we identify with the transition between the two critical phases of the $J_1$--$J_2$ chain. We then expand on the duality between the local dimer field and the instantons of $\hbm$, which is crucial to understanding the different phases. Finally, we present DMRG calculations which show an abrupt change in the behaviour of the lattice dimer correlations at $J_2^* \approx -1.3$.

\vspace{0.2cm}
\sectionn{One-loop RG flow}
If the dimer field is gapped, we can expand around $\chi = 0$, and the action \eqref{eq:J1J2_action} reduces to
\eqa{
\mS \sim i\Omega\bigr\rvert_{\chi=0} + \int &d\tau dx \Bigl[\gamma_{\hbm}(\pd_\mu \hbm)^2 + \gamma_\chi(\pd_\mu \chi)^2 \nn \\
&\;\;\;+ M^2 \chi^2 - g \chi^2 (\pd_\mu \hbm)^2 + \dots \Bigr],
\label{eq:Potts_phase_action}
}
for some renormalised couplings, where we have retained the lowest order coupling between the dimer and magnetic sectors. The WZ term simply reduces to the theta term (at $\theta = \pi$),
\eqn{
i\Omega\bigr\rvert_{\chi=0} = i\Theta(\pi).
\label{eq:Potts_phase_topological_term}
}

Now, the standard way of thinking about the RG flow of NLSMs is to use real space, block spin transformations; momentum space methods are confounded by the hard constraint $\hbm(x)^2 = 1$.
 
Our aim here, however, is simply to determine whether the phase where the dimer field is gapped is perturbatively stable; the N\'eel field $\hbm$ will be critical in either case, so we can simply write down its correlations in momentum space. This means, of course, that we are not proving that $O(3)_\pi$ renormalises to $\su(2)_1$~\footnote{There is no rigorous proof known that $O(3)_\pi$ is actually $\su(2)_1$ in the infrared, but it is widely expected to be.}, but it will allow us to self-consistently distinguish between the two phases.

The bare values of the couplings in Eq.~\eqref{eq:Potts_phase_action} are
\eqa{
&\gamma_{\hbm}(0) = \gamma_{\chi}(0) = g(0) = \gamma = \sqrt{\frac{1 - 2J_2}{32}}, \nn \\
\;\;\; &M^2(0) = -\frac{J_2}{4} > 0,
\label{eq:bare_values}
}
and the bare propagator for $\chi$ is simply
\eqn{
\begin{gathered}
\tik{
\draw (0, 0) -- (1.5, 0);
}
\end{gathered}
\;\;\;=\;\;\;
\mathcal{G}_{\chi}^{(0)}(p) \;=\; \frac{1}{\gamma_\chi p^2 + M^2}.
}
As mentioned above, we assume the correlations of $\hbm$ are critical, with some scaling dimension $\Delta > 0$. Thus
\eqa{
&\avg{\hbm(x)\cdot\hbm(x')} \sim \frac{1}{|x-x'|^{2\Delta}} \nn \\
&\Rightarrow \;\;\; \mathcal{G}_{\hbm}(p) = \avg{\hbm(p)\cdot\hbm(-p)} = \frac{\mC(\Lambda)}{|p|^{2 - 2\Delta}}.
}
Since we expect the correlations of $\hbm$ to follow from $\su(2)_1$, we set $\Delta = 1/2$ in what follows. We fix the overall factor $\mC(\Lambda)$, which depends on the ultraviolet cutoff $\Lambda$, by imposing the constraint $\hbm(x)^2 = 1$ on average. That is,
\eqn{
1 = \int\frac{d^2p}{(2\pi)^2}\mG_{\hbm}(p) = \mC\int_0^{\Lambda} \frac{d|p|}{2\pi} \;\;\Rightarrow\;\; \mC(\Lambda) = \frac{2\pi}{\Lambda}.
}
The field that appears in the action, however, is $\pd_\mu\hbm$, not $\hbm$ itself; its Green's function is, then,
\eqn{
\begin{gathered}
\tik{
\draw[dashed] (0, 0) -- (1.5, 0);
}
\end{gathered}
\;\;\;=\;\;\;
\mathcal{G}_{\pd \hbm}(p) \;=\; |p|^2 \mathcal{G}_{\hbm}(p) \;=\; \frac{2\pi|p|}{\Lambda}.
}

%
To obtain the flow equations we need the corrections to the propagator of the $\chi$ field,
\eqn{
\mG_{\chi}(p)
\;\;\;=\;\;\;
\begin{gathered}
\tik{
\draw (1, 0) -- (2.5, 0);
}
\end{gathered}
\begin{gathered}
\;\;\;+\;\;\;
\end{gathered}
\begin{gathered}
\tik{
\draw (1, 0) -- (2.5, 0);
\draw [dashed] (1.75, 0.5) circle [radius=0.5];
\fill[black] (1.75,0) circle (2pt);
}
\end{gathered}
\begin{gathered}
\;\;\;+\;\;\;
\end{gathered}
\begin{gathered}
\cdots,
\end{gathered}
}
which corrects the mass through the self-energy,
\eqn{
\Sigma\;\;\;=\;\;\;
\begin{gathered}
\tik{
\draw [dashed] (2, 0.5) circle [radius=0.5];
\fill[black] (2,0) circle (2pt);
}
\end{gathered}
\;\;\;+\;\;\;
\cdots,
}
and the vertex corrections, which are captured perturbatively by the expansion
\eqn{
\begin{gathered}
\tik{
\fill[gray] (0, 0) circle (4pt);
\draw (0,0) circle [radius=4pt];
}
\end{gathered}
\;\;\;=\;\;\;
\begin{gathered}
\tik{
\fill[black] (0, 0) circle (2pt);
}
\end{gathered}
\;\;\;+\;\;\;
\begin{gathered}
\tik{
\draw (0,0) arc (180:0:0.5);
\draw [dashed] (0,0) arc (-180:0:0.5);
\fill[black] (0, 0) circle (2pt);
\fill[black] (1, 0) circle (2pt);
}
\end{gathered}
\;\;\;+\;\;\;\cdots\;\;\;.
}

\begin{figure}[!t]
    \centering
    \includegraphics[]{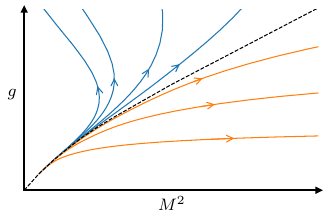}
    \caption{ 
    Flow diagram from the RG equations \eqref{eq:m2_flow_equation} \& \eqref{eq:g_flow_equation} of the decoupled action \eqref{eq:Potts_phase_action} (no numbers are shown because the choice of ultraviolet cutoff $\Lambda$ is arbitrary). Arrows indicate flow to the infrared. For sufficiently large bare values of $M^2$ we remain in the Zirnbauer phase where the dimer field is gapped (orange trajectories). As the bare value of $M^2$ is lowered, however, it will eventually flow to zero and the joint order parameter will recouple in the infrared (blue trajectories), leading us to the Affleck-Haldane phase (though the flow equations should not be trusted at small $M^2$). All the flows are away from the unstable $c = 2$ fixed point at $g = M^2 = 0$ (which is the direct sum of $O(3)_\pi$ with a free massless boson) in accordance with Zamolodchikov's $c$-theorem~\cite{zamolodchikov1986irreversibility}; though this unstable fixed point cannot be accessed by the spin chain as there is only one tuning parameter $J_2$.
    }
    \label{fig:flow_diagram}
\end{figure}

%
We begin with the mass. The lowest order term in the self-energy is given by
\eqn{
\begin{gathered}
\begin{tikzpicture}[baseline=6]
\draw [dashed] (2, 0.5) circle [radius=0.5];
\fill[black] (2,0) circle (2pt);
\end{tikzpicture}
\end{gathered}
\;\;=\;\;
-g\int \frac{d^2p}{(2\pi)^2} \mathcal{G}_{\pd\hbm}(p)
\;=\;
-\frac{g\Lambda^2}{3}.
\label{eq:m2_flow_equation}
}
This self-energy is subtracted from the effective mass term $M^2$, and so the flow equation at this order is
\eqn{
\beta(M^2) := -\Lambda\frac{dM^2}{d\Lambda} = 2M^2 - \frac{2\Lambda^2}{3}g,
}
where the first term corresponds to the fact that $M^2$ has engineering dimension 2.

We require also the vertex corrections. We again have only one diagram to consider, which we evaluate assuming zero external momentum,
\eqa{
\begin{gathered}
\tik{
\draw (0,0) arc (180:0:0.5);
\draw [dashed] (0,0) arc (-180:0:0.5);
\fill[black] (0, 0) circle (2pt);
\fill[black] (1, 0) circle (2pt);
}
\end{gathered}
&
\;\;\;=\;\;\;
(-g)^2 \int \frac{d^2p}{(2\pi)^2} \mathcal{G}_{\chi}^{(0)}(p) \mathcal{G}_{\pd\hbm}(-p)
\nn \\
\;=\;
&\frac{g^2}{\gamma_{\chi}}\left(1 - \frac{M\arctan\left(\frac{\gamma^{1/2}_{\chi}\Lambda}{M}\right)}{\gamma^{1/2}_{\chi}\Lambda}\right).
}
Noting that the bare vertex strength is $-g$ (but $g > 0$), and treating this purely perturbatively, we obtain the flow equation
\eqa{
\beta(g) := 
&\;\frac{g^2}{\gamma_{\chi}}\left(\frac{M\arctan\left(\frac{\gamma^{1/2}_{\chi}\Lambda}{M}\right)}{\gamma^{1/2}_{\chi}\Lambda} - \frac{M^2}{M^2 + \gamma_{\chi}\Lambda^2}\right) \nn \\
&\sim \frac{2\Lambda^2 g^2}{3M^2} + \mO\left(\frac{1}{M^4}\right).
\label{eq:g_flow_equation}
}
We observe that the strength of the vertex always increases towards the infrared, which suggests a generic instability for the joint order parameter to recouple -- however, this may be arrested by the flow of the mass $M^2$. We note that this flow is one-parameter renormalisable, since the beta function of the dimensionless ratio $\Lambda^2 g/M^2$,
\eqn{
\beta\left(\frac{\Lambda^2 g}{M^2}\right) \sim -\frac{4\Lambda^2 g}{M^2} + \frac{4\Lambda^4 g^2}{3M^4} + ...\;,
\label{eq:one_parameter_renormalisation}
}
is, to second order, a function only of $\Lambda^2 g/M^2$ itself.

%

A flow diagram obtained by numerically solving Eqs.\ \eqref{eq:m2_flow_equation} \& \eqref{eq:g_flow_equation} is shown in Fig.~\ref{fig:flow_diagram} (the full form of Eq.~\eqref{eq:g_flow_equation} is used). We observe that the flow equations predict a transition at a finite $M_*^2$ (and thus finite $J_2^*$), between the Affleck-Haldane phase with a joint-order parameter, and the Zirnbauer phase where the dimer component becomes gapped and decouples.

As mentioned, this corresponds to a transition between the two pictures of the RG flow to $\mathfrak{\hat{su}}(2)_1$ -- whereas Affleck and Haldane showed~\cite{affleck1987critical} that a perturbatively small mass term is irrelevant, we have shown that the phase where the dimer field is massive is stable for a sufficiently large (but finite) bare mass. We note that for $O(3)_\pi$ itself, where the bare mass is infinite, the present analysis is in favour of the Zirnbauer picture.

\vspace{0.2cm}
\sectionn{Duality between the dimer field and the instantons}
There is an important question, however, that we have not yet properly addressed: if the critical part of the theory, $\su(2)_1$, is the same in both phases, then both phases must have some field corresponding to $\tr(\hat{g})$ (cf. Eq.~\eqref{eq:matrix_field}) -- which must, therefore (i) be invariant under the global $SU(2)$ symmetry, (ii) be odd under translation by one site, and (iii) have power-law correlations.

In the Affleck-Haldane phase, this is clearly the local dimer field. But if $\chi$ is to appear as a separate gapped boson in the Zirnbauer phase, what are we to identify with $\tr(\hat{g})$?

The answer can be found by considering the expansion of the WZ term around $\chi = 0$ more carefully. We have
\eqn{
i\Omega(\chi)
\sim i\Theta(\pi) + \frac{i}{\pi}\int d\tau dx \,\chi\, \hbm\cdot(\pd_\tau\hbm\times\pd_x\hbm).
}
That is, the local dimer field $\chi$ is dual to the instanton density $\mathcal{I} := \hbm\cdot(\pd_\tau\hbm\times\pd_x\hbm)$. If we assume $\chi$ is massive (which is true, at least, in the ultraviolet) and integrate it out with the quadratic terms of Eq.~\eqref{eq:Potts_phase_action}, we obtain
\eqn{
\int d^2p\, \frac{\mathcal{I}_p \mathcal{I}_{-p}}{4\pi^2 (\gamma_{\chi}p^2 + M^2)} \sim \frac{1}{4\pi^2 M^2}\int d\tau dx\, \mathcal{I}^2.
}
In the Affleck-Haldane phase where ${M^2 \to 0}$, instantons of $\hbm$ are heavily suppressed (cf. the fact that ${\pi(S^2 \to S^3)}$ is trivial); but in the Zirnbauer phase, where instead the local dimer field $\chi$ is gapped and ${M^2 \to \infty}$, this suppression of the instantons is lifted.

As an aside, attempting to describe the spontaneous dimerisation transition at $J_2^d \approx +0.24$ with the spin coherent state path integral further bears out this duality -- in the $O(4)$ theory~\eqref{eq:J1J2_action}, $\chi$ simply acquires a non-zero expectation value; but in the $O(3)$ theory~\eqref{eq:O3NLSM} there is a `proliferation of instantons'~\cite{senthil2006competing}, which is typical of such dualities~\cite{francesco2012conformal,altland2010condensed}.

\vspace{0.2cm}
\sectionn{Density matrix renormalisation group calculations}
To check the predictions of the MPS field theory and the one-loop RG, we perform density matrix renormalisation group (DMRG) calculations on finite open-boundary chains. We use the ITensor package \cite{itensor,itensor-r0.3} to compute the ground state of Eq.~\eqref{eq:Hamiltonian} for various chain lengths up to $L = 1024$ and couplings $J_2$; the results are converged in the bond dimension. We perform DMRG sweeps until the energy density and half-chain entanglement entropy between sweeps are constant up to a precision of $10^{-9}$ and $10^{-7}$, respectively.

\begin{figure*}
    \centering
    \includegraphics[width=0.9\linewidth]{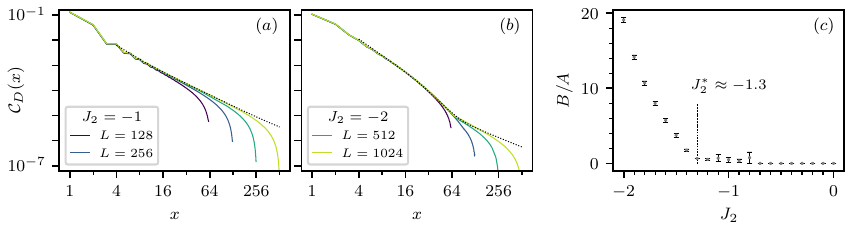}
    \caption{Transition between the Affleck-Haldane and Zirnbauer phases in the $J_1$--$J_2$ chain. (a), (b) Dimer correlations for $J_2 = -1$ and $-2$, respectively; dotted lines are the fits to Eq.~\eqref{eq:dimer_corr_fit}. We observe a region of exponential decay in the latter, but not the former. (c) Ratio $B/A$ (cf. Eq.~\eqref{eq:dimer_corr_fit}) of the exponential and power-law terms in the dimer correlators~\eqref{eq:dimer_corrs}; we find the exponential-decay term appears suddenly around $J_2^*\approx -1.3$. Error bars are obtained from the covariance matrix by Gaussian error propagation.
    }
    \label{fig:DMRG}
\end{figure*}

To get an estimate of where the transition occurs, we look at the correlations of the lattice dimer operator,
\eqa{
\hat{D}_x &= (-1)^x(\hS_x^+ \hS_{x+1}^- + \hS_x^- \hS_{x+1}^+ - \hS_{x-1}^+ \hS_{x}^- - \hS_{x-1}^- \hS_{x}^+), \nn \\
&\mC_{D}(x) = \langle\hat{D}_{L/2}\hat{D}_{L/2 + x}\rangle - \langle\hat{D}_{L/2}\rangle\langle\hat{D}_{L/2 + x}\rangle
\label{eq:dimer_corrs}
}
where we measure the correlations from the centre of the chain to minimise boundary effects. 
Now, we expect that the lattice operator $\hat{D}_x$ will overlap strongly with the local dimer field $\chi$ in the continuum -- with power-law correlations in the Affleck-Haldane phase, and exponentially-decaying correlations in the Zirnbauer phase. However, it is likely that $\hat{D}_x$ also has some small overlap with the emergent instanton field, and thence inherits power-law decay on the longest length-scales after initially decaying exponentially.

Nevertheless, if there is a distinguishable regime of exponential decay for $J_2 < J_2^*$, and no such regime for $J_2 > J_2^*$, that rather suggests that the local dimer field indeed acquires a mass for $J_2 < J_2^*$.
For each $J_2$, we fit
\eqa{
\mC_D(x) \approx \frac{A + B e^{-x/\xi}}{x \log(x)^{3/2}},
\label{eq:dimer_corr_fit}
}
where we include the logarithmic corrections~\cite{weber2023quantum,affleck1989critical,giamarchi1989correlation}. Note that we fix the maximum correlation length to $\xi_{\max} = 128$, to preclude the situation where the terms become indistinguishable. We show examples of the fits to the dimer correlations in Figs.~\ref{fig:DMRG}(a), (b); and we plot the ratio $B/A$ (cf. Eq.~\eqref{eq:dimer_corr_fit}) of the exponential and power-law terms in Fig.~\ref{fig:DMRG}(c), and find the exponential term emerges suddenly below $J_2^* \approx -1.3$.

\vspace{0.2cm}
\sectionn{Discussion and Conclusions}
In this Letter, we have shown that there is a critical-to-critical phase transition in the antiferromagnetic $J_1$--$J_2$ chain. The transition is driven by adding ferromagnetic second-neighbour couplings ${J_2 < 0}$ which do \textit{not} compete magnetically with the nearest-neighbour antiferromagnetic interactions ${J_1 > 0}$, but decouple the joint N\'eel-dimer order parameter by gapping the local dimer field; its dual field -- instantons of the N\'eel order -- is critical instead. We have shown the mass term for the dimer field is relevant below a finite $J_2^*/J_1 \approx -1.3$; the numerical value of the transition point is extracted from DMRG calculations, which reveal a sudden change in the nature of the lattice dimer correlations. 

We have shown how these phases distinguish two different pictures for the RG flow of $O(3)_\pi$ to $\su(2)_1$, and elucidated the physical differences between them, improving our understanding of one of the most well-studied and important field theories.
Further, the validity of the Zirnbauer picture may, according to Ref.~\cite{zirnbauer2024infrared}, have implications for the one-parameter renormalisability of NLSMs of other topological classes~\cite{heinzner2005symmetry}, including the Pruisken model of the integer quantum Hall transition~\cite{pruisken1984localization,zirnbauer1994towards,read2001exact,zirnbauer2019integer}. And furthermore, given the plethora of systems described by $O(3)_\pi$, or critical NLSMs more generally -- including higher half-odd-integer-$S$ spin chains, $SU(N)$ chains, the surfaces of topological insulators, Anderson transitions, and string worldsheets -- we suspect the $J_1$--$J_2$ spin chain will not host the only example of this or similar transitions.

In addition to finding a critical-to-critical transition in one of the canonical models of quantum magnetism, and providing strong evidence supporting the Zirnbauer picture for the RG flow of $O(3)_\pi$, these results serve to demonstrate the utility of the MPS path integral.
By including entanglement directly at the saddle-point level of the path integral, the MPS field theory is, perhaps, uniquely well-placed to sleuth out phase transitions involving a change in the entanglement structure.

We thank P. Fendley, M. Greiter, and F. Essler for helpful discussions and comments. This work was in part supported by NextGenerationEU under `Critical Properties of Quantum Ergodicity Breaking' (project id CN00000013), by the EPSRC under EP/S005021/1 and EP/I031014, and by the ERU under `Perspectives of a Quantum Digital Transformation'.  We acknowledge the IQTN (EP/W026872/1) for hosting fruitful discussions of the project. CAH is grateful for the hospitality of the Max Planck Institute for the Physics of Complex Systems (MPI-PKS) in Dresden, Germany, where part of this work was carried out.




\bibliography{refs_critical} 

\end{document}